\documentclass[final,5p,times,twocolumn]{elsarticle}
\usepackage{lineno}
\usepackage{amsfonts,physics}
\usepackage[utf8]{inputenc}
\usepackage{hyperref}
\hypersetup{colorlinks=true,urlcolor=red, citecolor=blue,linkcolor=blue}
\usepackage{amsmath}
\usepackage{amsfonts}
\usepackage{amssymb}
\usepackage{graphicx}
\begin{document}
	\title{Quantum phase properties of a state driven
		by a classical field}
	\author{Naveen Kumar}
	\ead{naveen74418@gmail.com}
	\author{Arpita Chatterjee$^*$}
\ead{arpita.sps@gmail.com}
\cortext[cor1]{Corresponding author}
\date{\today}
\address{Department of Mathematics, J. C. Bose University of Science and Technology,\\ YMCA, Faridabad 121006, India}
\begin{abstract}
	We consider a nonclassical state generated by an atom-cavity field interaction in presence
	of a driven
	field. In the scheme, the two-level atom is moved through the cavity and driven by a classical field. The atom interacts dispersively with the cavity field, which results in a  photon-number-dependent Stark shift. Assuming that the atom enters the cavity in the excited state $\ket{a}$, the obtained output cavity field is taken into account. The state vector $|\psi(t)\rangle$ describes the entire atom-field system but in our work we deal with the statistical aspects of the cavity field only. The quantum state that corresponds to the output cavity field is obtained by tracing out the atom part from $\ket{\psi(t)}\bra{\psi(t)}$. Different quantum phase properties such as quantum phase distribution, angular $Q$ phase function, phase dispersion are evaluated for the obtained radiation field. The second-order correlation function $g^2(0)$, an indirect phase characteristic is also considered.
\end{abstract}
\maketitle
\section{Introduction}
Nonclassical states (i.e. quantum states without any classical counterpart) are the states that can be characterized by the negative Glauber-Sudarshan $P$ function \cite{glauber,sudharshan}. The negative area of Wigner function also signifies the highly nonclassical character of a quantum state. In the last few years, several novel uses of nonclassical states have been described. For instance, squeezed states are employed in continuous-variable quantum cryptography \cite{hill}, most classical-like coherent state (coherent states are not entirely classical and do exhibit quantum characters) \cite{girish} is an appropriate candidate for quantum teleportation protocol \cite{furu}, and antibunching is demonstrated to be useful in the construction of single photon sources \cite{yuan}. Entangled states have emerged as one of the primary resources of quantum information processing as they proved to be necessary for implementing a set of protocols for discrete \cite{ekert} and continuous-variable quantum cryptography, quantum teleportation \cite{benn}, and dense coding \cite{hbenn} etc. These states of light are good resources for fundamental research in quantum mechanics and crucial for quantum information science in order to reach faster quantum processing, more secure quantum communication, and more accurate quantum metrology work. Here we wish to investigate the phase properties of such states.

Moreover, interest in applying open system theories in quantum information science has been developed due to the excellent success attained in the experimental front in manipulating quantum states of matter for quantum communication \cite{bouwm}. In the early days, the difficulty of writing a Hermitian operator for quantum phase was a well-known problem. Susskind-Glogower \cite{suss} and Pegg-Barnett \cite{pegg} introduced various methods to express quantum phase as a Hermitian operator. In a finite-dimensional Hilbert space, Pegg and Barnett \cite{pegg2,pegg3} following the efforts of Dirac \cite{direc}, performed a polar decomposition of the annihilation operator and defined a Hermitian phase operator. Their method involves first performing the expectation value of a function of the phase operator in a finite-dimensional Hilbert space before increasing the dimension up to the limit of infinity. This expectation value, however, cannot be understood as a function of a Hermitian phase operator in an infinite-dimensional Hilbert space. Dirac, one of the inventors of quantum mechanics, was credited for developing the polar decomposition method for introducing the unitary phase operator $U=\exp(i\theta)$. Unfortunately, the concept of the Dirac phase operator is found unacceptable since it was unable to explain the uncertainty relation and the Hermitian structure of the phase operator. Louisell \cite{lsl} defined a meaningful phase operator in terms of periodic functions and resolved those difficulties. In recent years, we have seen different approaches to describe phase functions followed by multiple applications of quantum phase distribution and quantum phase fluctuations. It is important to note that the idea of phase is incredibly helpful for explaining a variety of quantum optical phenomena, namely quantum random number generation \cite{xu,raff}, cryptanalysis of squeezed state-based continuous-variable quantum cryptography \cite{horak}, generation of solitons in a Bose-Einstein condensate \cite{den}, storing and retrieving information from Rydberg atoms \cite{ahn, gisin}, phase imaging of cells and tissues for biomedical applications \cite{park}, and figuring out the transition temperature for superconductors \cite{emery}.

Phase diffusion was investigated previously in the context of quantum phase dispersion under environmental effects \cite{ban1,ban2,abdel,ban3}. Quantum phase distribution has also been suggested as a phase fluctuations measurement known as phase dispersion \cite{ban1}. The standard deviation of an observable is normally regarded as the most natural measure of quantum fluctuations associated with that observable \cite{orlo}, and quantum fluctuation reduction below the coherent state level corresponds to a nonclassical state. However, standard deviations can be merged to generate more complex measurements of nonclassicality, which may increase as nonclassicality increases. Reducing the Carruthers-Nieto symmetric quantum phase fluctuation parameter $U$ \cite{nito} with respect to its Poissonian state value leads to an antibunched state, while the reverse is not true \cite{gupta}. As a result, decrease of $U$ is a stronger criterion of nonclassicality than lower-order antibunching.

Studying quantum properties of micro- or nano-mechanical systems has developed notable interest within the physics community in the last decade \cite{joshi1,nasir1,nasir2,anes}. This is due to the emerging feasibility of ultra-cooled mechanical systems which were impossible before
the advent of laser cooling. As per the prevailing conditions, these nanomechanical systems can be fabricated. Because of their complex nature, it is not always possible to analyze these setups theoretically. We therefore investigate a less complex nanomechanical structure which is closely related to various physical systems such as a trapped ion or a superconducting qubit interacting with a quantized electromagnetic mode. In the single excitation limit, the interaction between a two-level atom and a single-mode cavity field is analogous to the interaction between a quantum harmonic oscillator and a cavity field which is often used to study fundamental quantum phenomena and serves as a basis for many light-matter experiments at the quantum level. Another key feature of this atom-cavity interaction at a higher level includes creation of entangled states, where the quantum states of the atom and the field become correlated in a nonclassical way. This concept is crucial for understanding quantum information processing and quantum communication. The interaction of the system with the environment, such as the surrounding electromagnetic field, can lead to decoherence and thus deteriorates the quantum properties of the atom-field system. This process has implications for quantum measurement and the maintenance of quantum coherence in quantum information processing. On the other hand, researchers can use external fields to manipulate the atom-field state and that control over the quantum state is a cornerstone of quantum technologies, including quantum sensors and quantum-enhanced metrology. This
model is also useful for further investigating quantum phase properties like phase distribution, phase fluctuation, angular $Q$ function, phase dispersion and indirectly photon antibunching
of various radiation fields. These phase properties are very important as they influence a wide range of physical phenomena and applications. The phase distribution of an atom-cavity system describes the structure from an interferometric point of view and determines the possible interference patterns. The well-defined phase relationships lead to coherent interference, which is central to various experiments involving beam-splitters, interferometers, and quantum interference devices. In experimental framework, maintaining a stable phase relationship is primarily required. Phase stabilization techniques are used to ensure the coherence of quantum states and to minimize phase fluctuations that can lead to errors in measurements and quantum operations \cite{xu,raff}. In quantum state tomography, the idea is to characterize and reconstruct the complete quantum state of a system \cite{horak,den,ahn,gisin}. Phase information provides insights into the structure of the quantum state that is complementary to that of the output intensity \cite{park}. Phase dispersion refers to the phenomenon in which different wavelengths or frequencies of light travel at different speeds through a medium, which causes a distortion or spreading of the wavefront. This can affect the overall behavior of the light as it propagates through the cavity medium \cite{emery}. This is significant for phenomena like dispersion-induced transparency and group velocity control. In addition, the present model of atom-cavity system could be beneficial for inspecting quantum state transfer \cite{joshi2,joshi3}
which appears, for example, in the context of a chain of coupled harmonic oscillators. Hence this paper aims to discuss the phase properties of the cavity field in an atom-cavity system coupled with the atom being driven by a classical field. The influence of different state parameters on the quantum phase properties are to be investigated.

Understanding and controlling nonclassical states of light is one of the most fundamental objectives in the field of quantum optics \cite{arpita1}. It has been performed across a wide variety of physical systems ranging from atomic cavity quantum electrodynamics (QED) systems and nonlinear optical media to recent experiments with superconducting circuits \cite{lem}. A common way for characterizing the quantum nature of the nonclassical states is to quantify the intensity fluctuations via the correlation function $g^{(2)}(0)$, defined as $g^{(2)}(0)=\frac{\langle{:{\hat{I}}^2:}\rangle}{{\langle{\hat{I}}\rangle}^2}$, where $\hat{I}$ is the field intensity and colons indicate normal ordering. While classical intensity fluctuations always obey $g^{(2)}(0)\geq 1$, quantum states
can violate this bound. Therefore $g^{(2)}(0)< 1$ is often used as a criterion to identify nonclassical states. Antibunched photonic states are a type of nonclassical light that exhibit a reduced probability for two or more photons which are being detected simultaneously as compared to what is expected for classical light. One method for generating antibunched states is using cavity-QED systems \cite{wm}, that involves coupling of a single quantum emitter, such as an atom or a quantum dot, to a high-$Q$ optical cavity. The strong coupling between the emitter and the cavity can lead to the emission of single photons in a controlled manner, resulting in antibunched states. There may be other scenarios where the $g^{(2)}$ function shows antibunching. A trivial example is having a very weak but coherent field, which is not a signature of nonclassical state but of a very low driving intensity. In the low driving regime, antibunching results due to the so called photon blockade caused by the Rabi splitting of the originally harmonic spectrum. The second-order intensity correlation function $g^2(0)$, an antibunching measure \cite{g2} quantifies the probability of detecting two photons simultaneously. Mandel's $Q$ parameter provides a more general measure of photon number statistics. The nonclassicality of the considered state is reported in terms of the intensity correlation $g^2(0)$ that can be derived as a consequence of the generalized expectation values obtained in \ref{subsec2}.

We have organised this paper in the following manner. In Sec.~\ref{sec2}, a state is generated by an atom-field interaction in the presence of atomic driving. In Sec.~\ref{sec3}, we have studied the behaviour of different phase properties like quantum phase distribution, angular $Q$ phase function and phase fluctuation corresponding to our state of interest. Sec.~\ref{sec4} concludes the article.

\section{State of interest}
\label{sec2}
In the present scheme, a two-level atom is
sent through the cavity and driven by a classical field. 
The atom interacts dispersively with the cavity field, which induces photon-number dependent Stark shift. The Stark shift is a change in the energy levels of an atom caused by an external electric field. In this case, the interaction between the atom and the cavity field leads to a Stark shift that depends on the number of photons present in the cavity. This means that the magnitude of the Stark shift experienced by the atom is directly influenced by the number of photons in the cavity. 
The Hamiltonian describing the interaction between a weakly driven two-level atom and a single-mode cavity field reads as \cite{alsing,biao,arpita} (with unit Planck's constant $\hbar$  = 1)
\begin{eqnarray}
	\nonumber
H & = & H_\textrm{atom}+H_\textrm{cavity}+H_\textrm{interaction}
\\\nonumber
& = & \omega_0 S_z + \omega_a a^\dag a + g(a^\dag S^- + aS^+)\\
&+&
\epsilon (S^+ e^{-i\omega t} + S^- e^{i\omega t}),
\end{eqnarray}
where $a$, $a^\dagger$ are the bosonic annihilation and creation operators for the cavity mode. In addition, $S_z=\frac{1}{2}\big[\ket{a}\bra{a}-\ket{b}\bra{b}\big]$, $S^+=\ket{a}\bra{b}$, and $S^-=\ket{b}\bra{a}$ are the atomic spin operators for inversion, raising, and lowering, respectively, $\omega_0$, $\omega_a$, and $\omega$ are the frequencies for the atomic transition (between $\ket{a}$ and $\ket{b})$, for cavity mode, and for classical field, respectively. Here $g$ ($\epsilon$) is the coupling constant between the atom and cavity field (driving field).
If the system's reference frame is rotated with respect to the driving field frequency $\omega$, the Hamiltonian reduces to
\begin{eqnarray}
\label{eq2}
H_f & = & i\dot{R}(t)R^\dagger(t)+R(t)H(t)R^\dagger(t) \\\nonumber
& = & \Delta S_z+\delta a^\dag a+g(a^\dag
S^-+aS^+)+\epsilon(S^++S^-),
\end{eqnarray}
where $R(t)=e^{-i\omega t\,(S_z+a^\dagger a)}$ is the rotation operator and $\Delta = \omega_0-\omega$, $\delta = \omega_a-\omega$. The higher order terms in the exponent of \eqref{eq2} can be computed by using the well-known Baker-Campbell-Hausdorff (BCH) formula as follows:
\begin{eqnarray*}
e^{A}Be^{-A}=B+\frac{1}{1!}[A,B]+\frac{1}{2!}[A,[A,B]]+\ldots,
\end{eqnarray*}		
where $A$, $B$ are the operators. The classical field frequency is chosen in such an appropriate way so that the atomic
transition is tuned in resonance with the classical field. That means assuming $\Delta=0$, in the interaction picture the Hamiltonian is further simplified by using 
\begin{eqnarray*}
H_I & = & e^{iH_{f_0}t}H_{f_1} e^{-iH_{f_0}t}\\
& = & \frac{1}{2}g\left[|+\rangle\langle+|-|-\rangle\langle-|+e^{2i\epsilon t}|+\rangle\langle-|\right.\\
& & 
\left.- e^{-2i\epsilon t}|-\rangle\langle+|\right] ae^{-i\delta t}+\mathrm{H.c.,}
\end{eqnarray*}
where
\begin{eqnarray*}
H_{f_0} & = & \delta a^\dag a+\epsilon(S^++S^-),\\
H_{f_1} & = & g(a^\dag S^-+aS^\dag),
\end{eqnarray*}
where the dressed states $|\pm\rangle = \frac{1}{\sqrt{2}}(|b\rangle\pm|a\rangle)$
are the eigenstates of $(S^++S^-)$ with the eigenvalues $\pm 1$, $\ket{a}$ and $\ket{b}$ being the excited and ground states of the atom.
In the strong driving regime $\epsilon\gg\{\delta, g\}$, the effective Hamiltonian under the rotating-wave approximation (neglecting the highly oscillating terms like $a^\dagger S^+ $ and $a S^-$) is obtained as \cite{chen}
\begin{align}
	\nonumber
	H_{\mathrm{eff}}&= \frac{1}{2}g\left[\ket{+}\bra{+}-\ket{-}\bra{-} \right](ae^{-i\delta t}+a^\dag e^{i\delta t})\\\nonumber
	\label{heff}
	& = \frac{1}{2}g(S^++S^-)(ae^{-i\delta t}+a^\dag e^{i\delta t})\\
	& \approx \frac{1}{2}g (aS^+e^{-i\delta t}+a^\dag S^- e^{i\delta t})
\end{align}
We now proceed to solve the time-dependent Schr\"{o}dinger equation \cite{motion}
\begin{eqnarray}
\label{eq3}
i\hbar\frac{\partial |\psi(t)\rangle }{\partial t}= H_{\mathrm{eff}} |\psi (t)\rangle,
\end{eqnarray}
for the state vector $|\psi(t)\rangle = \sum_{n}^\infty\left[c_{a, n}(t) |a, n\rangle +c_{b, n}(t) |b,n\rangle\right]$, where $c_{a,n}(t)$, $c_{b,n}(t)$ are the probability amplitudes. At any time $t$, the state vector $\ket{\psi(t)}$ is the linear combination of the states $\ket{a,n}$ and $\ket{b,n}$. Here $\ket{a,n}$ is the state in which the atom in its excited state $\ket a$ and field has $n$ photons. A similar description exists for the state $\ket{b,n}$. The interaction energy (\ref{heff}) can only cause transitions between the states  $\ket{a,n}$ and $\ket{b,n+1}$. We therefore consider the evolution of the amplitudes $c_{a,n}$ and $c_{b,n+1}$. The equations of motion thus obtained for the probability amplitudes $c_{a,n}(t)$, $c_{b,n+1}(t)$ are
\begin{eqnarray}
\label{neweq1}
\dot{c}_{a,n}(t) & = & -\frac{ig\sqrt{n+1}}{2}  e^{i\delta t} c_{b,n+1}(t),\\
\label{neweq2}
\dot{c}_{b,n+1}(t) & = & -\frac{ig\sqrt{n+1}}{2}  e^{-i\delta t} c_{a,n}(t)
\end{eqnarray}
Differentiating  $\dot{c}_{a,n}(t)$ in \eqref{neweq1} with respect to $t$ one time and substituting $\dot{c}_{b,n+1}(t)$ from \eqref{neweq2}, we obtain a second-order differential equation as

\begin{eqnarray}
\ddot{c}_{a,n}(t)-i\delta\dot{c}_{a,n}(t)+\frac{g^2(n+1)}{4}{c}_{a,n}(t)=0
\end{eqnarray}

Different atom-cavity entangled states can be produced depending on the initial assumptions. Assuming that the atom enters the cavity in the excited state $|a\rangle$, $c_{a,n}(0) = c_n(0)$ and $c_{b,n+1}(0)=0$. A general solution of \eqref{eq3} with these conditions is
\begin{eqnarray*}
c_{a,n}(t) & = & c_{n}(0)\left\lbrace \cos{\frac{\Omega_{n}t}{2}}-\frac{i\delta}{\Omega_{n}} \sin{\frac{\Omega_{n}t}{2}}\right\rbrace e^{\frac{i\delta t}{2}}\\\\
c_{b,n+1}(t) & = & -c_{n}(0)  \frac{ig\sqrt{n+1}}{\Omega_{n}} \sin{\frac{\Omega_{n}t}{2}} e^{-\frac{i\delta t}{2}},
\end{eqnarray*}
where ${\Omega_n}^2=\delta^2 +g^2(n+1)$ and $c_n(0)$ is the field's amplitude only \cite{peng1}. If the field is initially in a coherent state then
$c_{n}(0)=e^{-\frac{|\alpha|^2}{2}}\frac{\alpha^n}{\sqrt{n!}}$. The state vector $|\psi(t)\rangle$ describes the entire atom-field entangled system but in our work, we deal with the statistical aspects of the cavity field only. The output state vector for the cavity field is obtained by tracing out the atom part from $\ket{\psi(t)}\bra{\psi(t)}$ as following:
\begin{eqnarray}
\label{eq10}
\ket{\psi_f}\bra{\psi_f} = \textrm{Tr}_{\textrm{atom}}\left[\ket{\psi(t)}\bra{\psi(t)}\right],
\end{eqnarray}
$\ket{\psi_f}$ is the state of interest in rest of the paper.

\subsection{Resonance case}
In case of $\delta=0$, ${\Omega_n}^2=g^2(n+1)$ and the probability amplitudes reduce to
\begin{eqnarray}
\begin{array}{lcl}
	c_{a,n}(t) & = & c_{n}(0) \cos{\left(\frac{\Omega_{n}t}{2}\right)}\\
&=&e^{-\frac{|\alpha|^2}{2}}\frac{\alpha^n}{\sqrt{n!}}\cos{\left(\frac{gt\sqrt{n+1}}{2}\right)}
\end{array}
\end{eqnarray}
and
\begin{eqnarray}
\begin{array}{lcl}
c_{b,n+1}(t) & = & -c_{n}(0)  \frac{ig\sqrt{n+1}}{\sqrt{g^2(n+1)}} \sin{\left(\frac{gt\sqrt{n+1}}{2}\right)}\\
&=& - i e^{-\frac{|\alpha|^2}{2}}\frac{\alpha^n}{\sqrt{n!}}\sin{\left(\frac{gt\sqrt{n+1}}{2}\right)}
\end{array}
\end{eqnarray}

\subsection{Generalized expectations}
\label{subsec2}
Different expectations with respect to the cavity field can be obtained as:

\begin{align*}
\bra{\psi_f}aa^\dag\ket{\psi_f}
&=\sum_{n=0}^\infty\big[(n+1)c_{a,n}(c^*_{a,n}+c^*_{b,n})\\
&+(n+2)c_{b,n+1}(c^*_{a,n+1}
+c^*_{b,n+1})\big]
\end{align*}
\begin{align*}
\bra{\psi_f}a^{\dag p}a^q\ket{\psi_f}&=
\sum_{n=0}^\infty\bigg[c_{a,n+p-q}^*c_{a,n}\sqrt{\frac{(n+p-q)!n!}{(n-q)!^2}} \\
&+ c^*_{a,n+p-q+1}c_{b,n+1}\sqrt{\frac{(n+p-q+1)!(n+1)!}{(n+1-q)!^2}}\\ 
&+c^*_{b,n+p-q}c_{a,n}\sqrt{\frac{(n+p-q)!n!}{(n-q)!^2}}\\
&+c^*_{b,n+p-q+1}c_{b,n+1}\sqrt{\frac{(n+p-q+1)!(n+1)!}{(n+1-q)!^2}}\bigg]
\end{align*}

\section{Quantum phase properties}
\label{sec3}
\subsection{Quantum phase distribution}
\begin{figure}[h]
\centering
\includegraphics[width=0.3\textwidth]{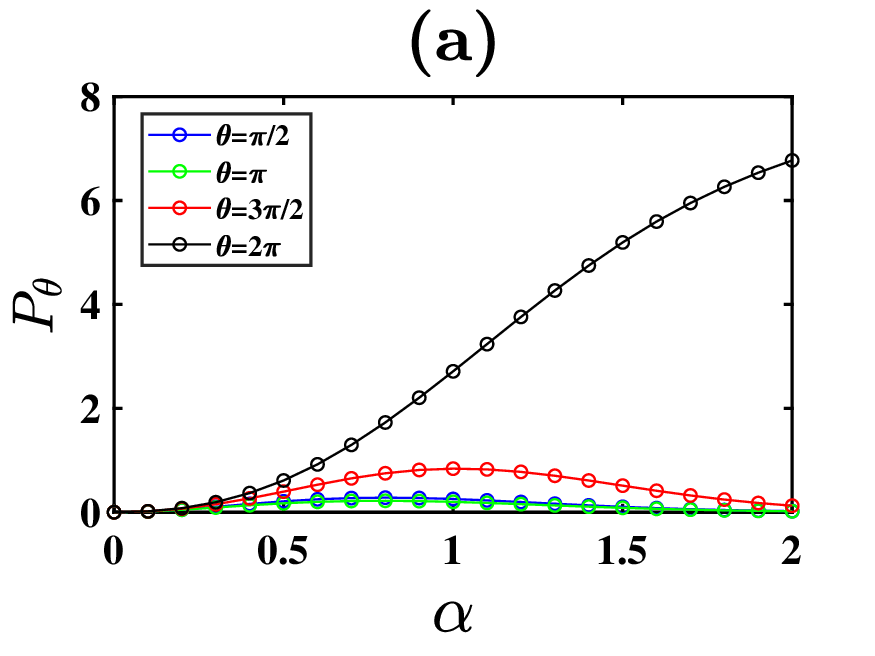}
\includegraphics[width=0.3\textwidth]{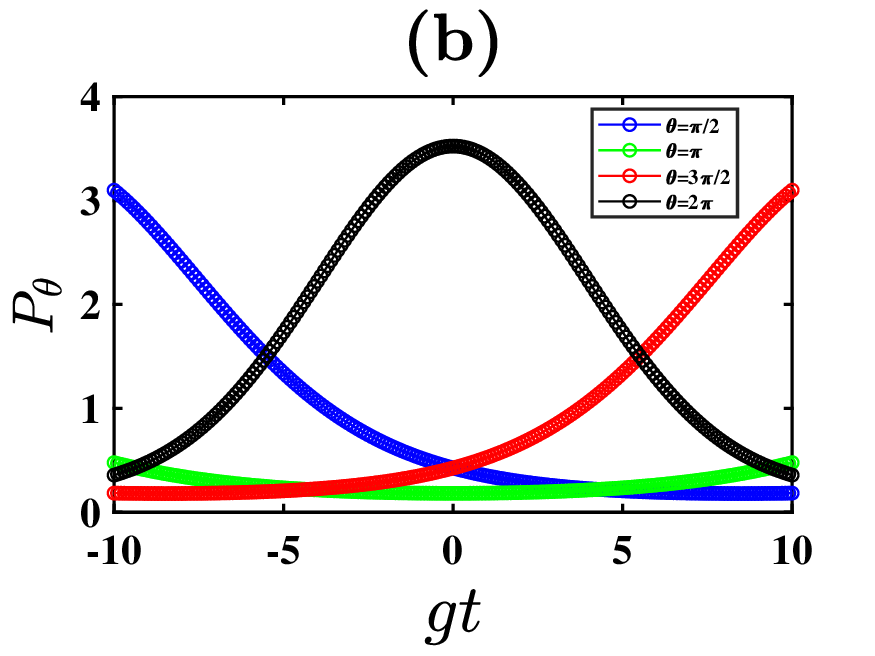}
\includegraphics[width=0.3\textwidth]{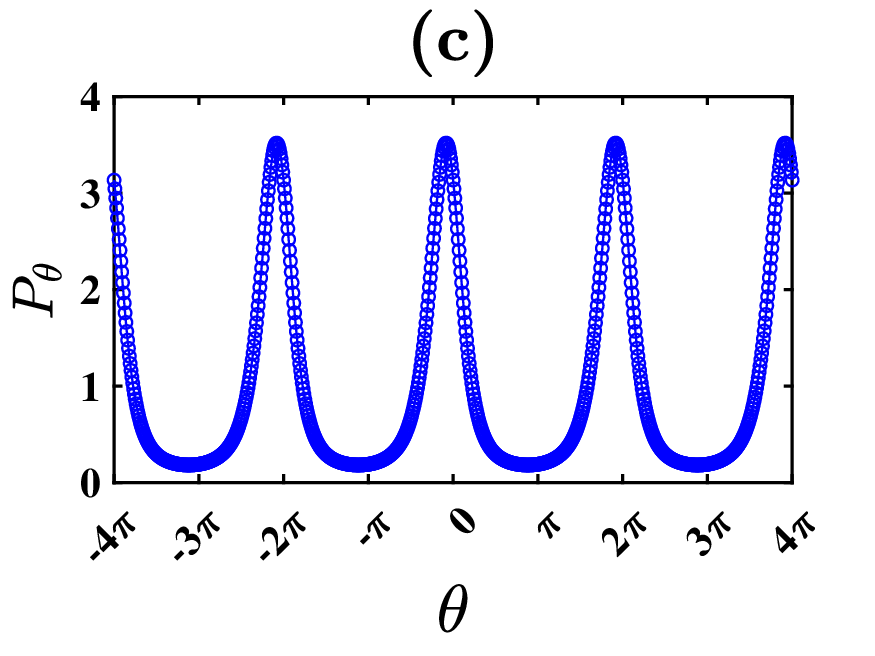}
\caption{Plot of quantum phase distribution function $P_{\theta}$ with respect to (a) $\alpha$ with $gt=3$ and different values of $\theta$, (b) $gt$ with $\alpha=1$ and different values of $\theta$, (c) $\theta$ with $gt=2$, $\alpha=1$.}
\label{qpd}
\end{figure}

A distribution function allows us to calculate the expectation values of an operator with respect to the corresponding
density matrix. The phase distribution function of a given density operator system is defined as \cite{ban1,tara}
\begin{align}
P_{\theta}&=\frac{1}{2\pi}\bra{\theta}\rho\ket{\theta},
\end{align}
where the phase state $\ket{\theta}$ in terms of the number state $\ket{n}$ is given by \cite{tara}
\begin{equation}
\ket{\theta}=\sum_{n=0}^{\infty}e^{in\theta}\ket{n}
\end{equation}
The analytical expression for the phase distribution function of the cavity field $\ket{\psi_f}\bra{\psi_f}$ is
\begin{equation}
\label{ptheta}
P_{\theta}=\frac{1}{2\pi}\left|\sum_{m=0}^{\infty}e^{im\theta}({c^*_{a,m}+c^*_{b,m+1}})\right|^2
\end{equation}
The phase distribution function $P_\theta$ is plotted as a function of $\alpha$ considering $gt=3$, as a function of $gt$ with $\alpha=1$ and as a function of $\theta$ substituting $gt=2$, $\alpha=1$ in Fig.~\ref{qpd}. 

While $P_\theta$ is plotted as a function of $\alpha$, it is observed that $P_\theta$ first increases then decreases as $\alpha$ changes from $0$ to $2$. With different increasing values of $\theta$ ($\pi/2$, $\pi$, $3\pi/2$ and $2\pi)$, the curves are ascending for greater values of $\alpha$. Here $P_\theta$ attains its maximum value for $\theta=2\pi$. For a fixed value of $gt=3$, the maximum value of $P_{\theta}$ is 0.22 for $\theta=\pi/2$ and $\pi$, 0.83 for $\theta=3\pi/2$, and 6.77 for $\theta=2\pi$ (see Fig.~1{\color{blue}(a)}). The plots of $P_{\theta}$ as a function of $gt$ are exhibiting oscillatory nature, and while $\theta=2\pi$, it is seen that $P_\theta$ is symmetric at $gt=0$ and attains the maximum value 3.5 (see Fig.~\ref{qpd}{\color{blue}(b)}). In Fig.~\ref{qpd}{\color{blue}(c)}, $P_{\theta}$ as a function of $\theta$ exhibits wave nature with varying amplitude.

\subsection{Quantum phase fluctuation}
To get rid of the limitations of the Hermitian phase operator structure of Dirac, Louisell first mentioned that bare phase
operator should be replaced by periodic functions. As a consequence, Susskind and Glogower \cite{suss} prescribed explicit form of phase operators involving
sine ($\hat{S}$) and cosine ($\hat{C}$) functions which are further modified by Barnett and Pegg \cite{pegg} as following:
\begin{equation}
\hat{C}=\frac{a+a^\dag}{2({\bar{N}+\frac{1}{2})}^{\frac{1}{2}}}
\end{equation}
and
\begin{equation} \hat{S}=\frac{a-a^\dag}{2i({\bar{N}+\frac{1}{2})}^{\frac{1}{2}}}
\end{equation}
where $\bar{N}$ is the average number of photons in the measured field. The quantum phase fluctuation parameters in terms of sine and
cosine operators are given as \cite{caru}
\begin{eqnarray}
U =\frac{{(\Delta N)}^2\left[{(\Delta S)}^2+{(\Delta C)}^2\right]}{\left[\langle \hat{S} \rangle^2 +\langle \hat{C} \rangle^2  \right]}
\end{eqnarray}
\begin{eqnarray}
S={(\Delta N)}^2{(\Delta S)}^2
\end{eqnarray}
and
\begin{eqnarray}
Q'=\frac{S}{{\langle\hat{C}\rangle}^2}
\end{eqnarray}
The phase fluctuations can be obtained by using the following expressions where the angle brackets denote quantum expectation values
\begin{eqnarray*}
\langle\hat{C}\rangle =\frac{1}{2({\bar{N}+\frac{1}{2})}^{\frac{1}{2}}}\left(\langle a\rangle+\langle a^\dag\rangle\right)
\end{eqnarray*}

\begin{eqnarray*}
\langle\hat{S}\rangle
=\frac{1}{2i({\bar{N}+\frac{1}{2})}^{\frac{1}{2}}}\left(\langle a\rangle-\langle a^\dag\rangle\right)
\end{eqnarray*}
\begin{eqnarray*}
\langle\hat{C}^2\rangle
=\frac{1}{-4({\bar{N}+\frac{1}{2})}}\left(\langle a^2\rangle+\langle a^{\dag 2} \rangle+\langle aa^\dag \rangle+\langle a^\dag a\rangle\right)
\end{eqnarray*}
\begin{eqnarray*}
\langle\hat{S}^2\rangle
=\frac{1}{-4({\bar{N}+\frac{1}{2})}}\left(\langle a^2\rangle+\langle a^{\dag 2} \rangle-\langle aa^\dag \rangle-\langle a^\dag a\rangle\right)
\end{eqnarray*}
and
\begin{align*}
\bar{N}&=\bra{\psi_f}N\ket{\psi_f}=\bra{\psi_f}a^\dagger a\ket{\psi_f}\\
\end{align*}
In recent days, a group of people provides a physical meaning to one of these fluctuation parameters by establishing its relation with
anti-bunching and sub-Poissonian photon statistics \cite{gupta}. Therefore, the quantum phase fluctuation investigated here with three parameters can also be used to identify the nonclassical character of the quantum states under investigation. In the Barnett-Pegg formalism, $U$ parameter proved itself relevant as a nonclassicality witness. Specifically $U$ is 0.5 for the coherent state, and a value of $U$ below the coherent state limit indicates the presence of nonclassical behavior \cite{gupta} in the considered quantum state. Here we see the variation of quantum phase fluctuation using three parameters $P$ ($S$, $Q'$ or $U$) to witness the nonclassical nature of the
quantum states under consideration. 

\begin{figure}[h]
\centering
\includegraphics[scale=.45]{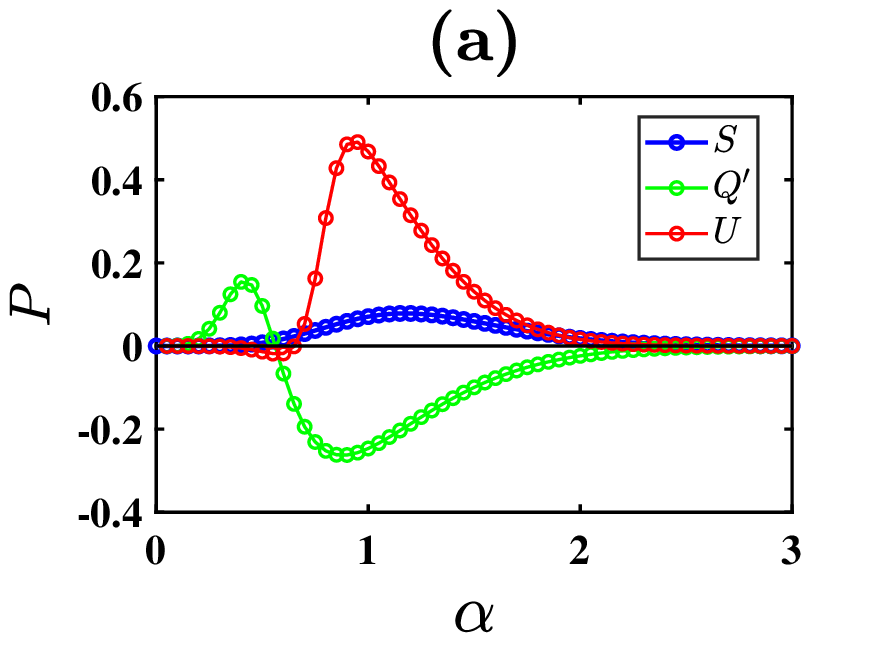}
\includegraphics[scale=.45]{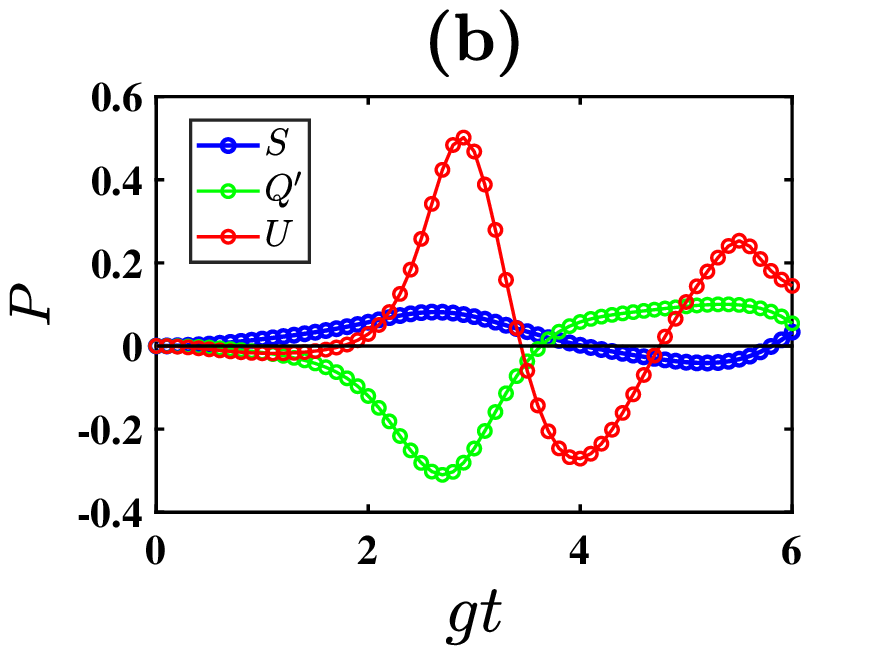}
\caption{Variation of phase fluctuation parameters $P=(S,Q',U)$ as a function of (a) $\alpha$ with $gt=3$, (b) $gt$ with $\alpha=1$.}
\label{squ}
\end{figure} 
The fluctuation parameters $P$ ($S$, $Q'$ or $U$) are plotted against the coherent state parameter $\alpha$ and the scaled time $gt$ in Fig.~\ref{squ}. It is found that two of the phase fluctuation parameters, namely  $U$ and $Q'$ become negative for some values of $\alpha$. These two parameters change in a zigzag way with the value of displacement parameter in the range $0\leq\alpha\leq2$ but the variation in the diagram of $S$ is less. On the other hand, while plotting against $gt$, parameter $S$ goes below zero as $4\leq gt\leq 6$, $Q'$ shows negative values whenever $1\leq gt\leq 3.6$. Also, $U$ becomes negative when $3.5\leq gt\leq 4.8$, which clearly demonstrates the nonclassical character of the cavity field state.

\subsection{Angular $Q$ function}
Another description of phase distribution defined as radius integrated quasidistribution function can be used as a witnesses for quantumness \cite{furu}. The angular $Q$ function, based on the angular part of the $Q$ function, is defined as
$$Q_{\theta_1}=\int_{0}^{\infty}Q(\beta,\beta^*)\,|\beta|\,d|\beta|$$
where the $Q$ function \cite{hsumi} is the projection of the state of interest on the coherent state basis, that is,
$$Q=\frac{1}{\pi}\bra{\beta}\rho\ket{\beta}$$
with the coherent state amplitude ${\beta}=|\beta|\,e^{i\theta_1}$. It has been extensively studied how useful the $Q$ function is in state tomography \cite{banerjee.A} and as a witness of nonclassicality \cite{furu}. Using the coherent state expansion $\ket{\beta}=e^{-{|\beta|^2}/{2}}\sum_{m=0}^{\infty}\frac{\beta^{*m}}{\sqrt{m!}}\ket{m}$, $Q$ function can be calculated as
\begin{align}	Q&=\frac{1}{\pi}\sum_{m,n=0}^{\infty}e^{-|\beta|^2}\frac{\beta^{*m}}{\sqrt{m!}}
\frac{\beta^{n}}{\sqrt{n!}}\left(c^*_{a,m}+c^*_{b,m+1}\right)\left(c_{a,n}+c_{b,n+1}\right)
\label{Q1}
\end{align}
Finally the angular $Q$ function can be obtained as
\begin{align}\nonumber
Q_{\theta_1}&=\int_{0}^{\infty}Q(\beta,\,\beta^*)|\beta|d|\beta|\\
& = \frac{1}{2\pi}\sum_{m,n=0}^{\infty}\frac{e^{i\theta_1(n-m)}}{\sqrt{m!n!}}\left(c^*_{a,m}+c^*_{b,m+1}\right)\left(c_{a,n}+c_{b,n+1}\right)
\Gamma\left(\frac{m}{2}+\frac{n}{2}+1\right),	
\end{align}
where $\Gamma(n)=\int_{0}^{\infty} e^{-x} x^{n-1} dx$ is the usual Gamma function.

\begin{figure}[htb]
\centering
\includegraphics[width=0.3\textwidth]{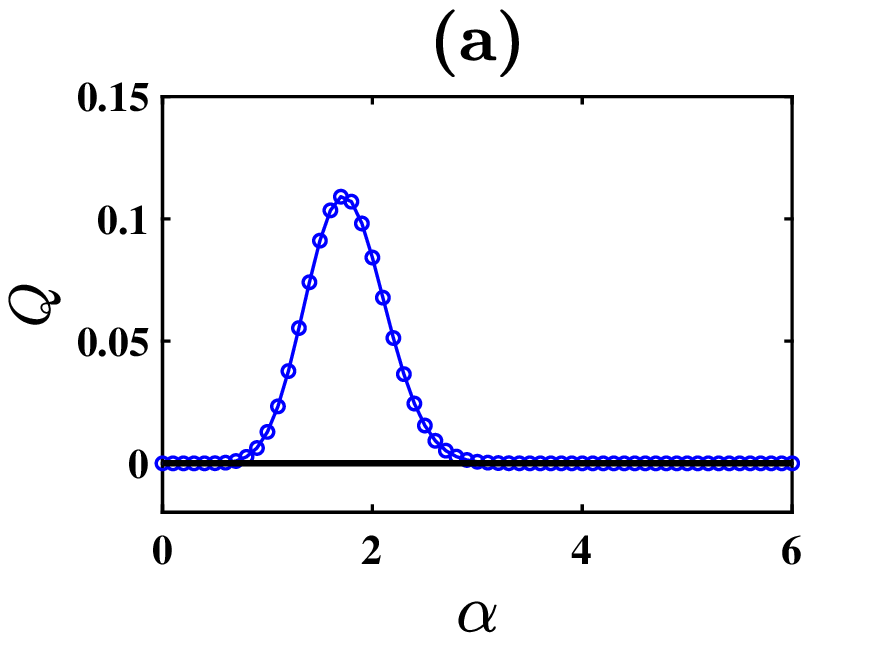}
\includegraphics[width=0.3\textwidth]{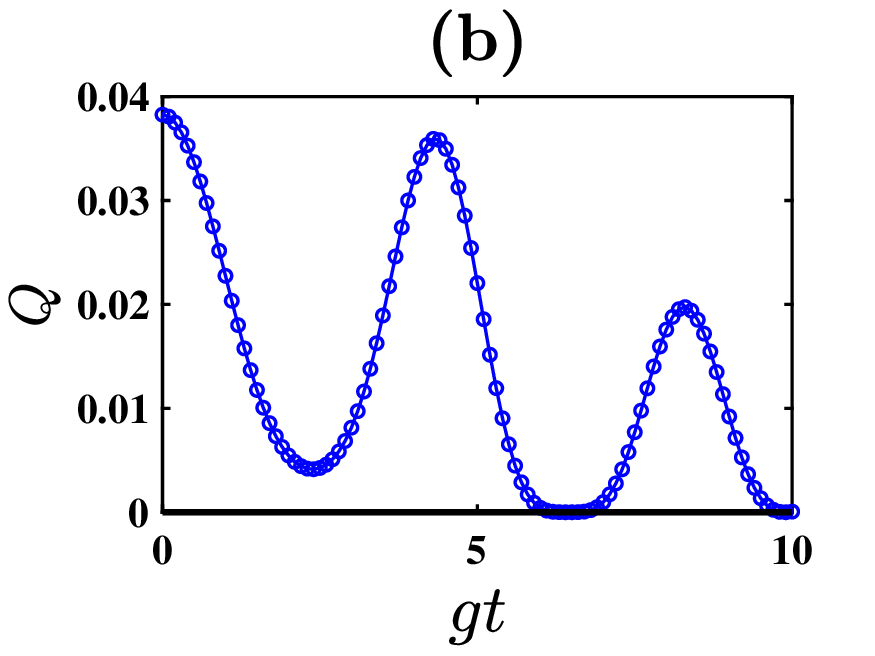}
\includegraphics[width=0.3\textwidth]{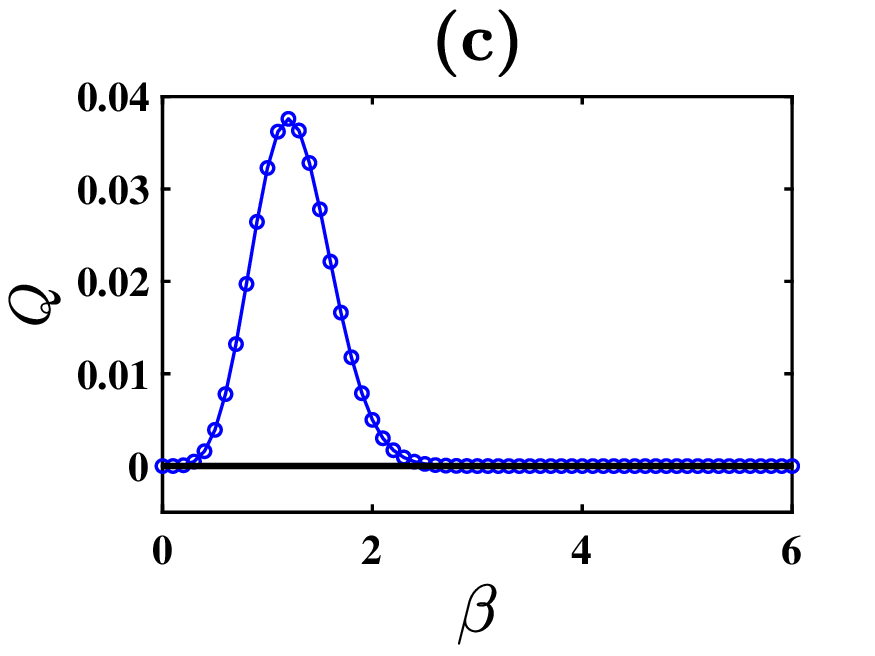}
\caption{Variation of $Q$ function with respect to (a) $\alpha$ with $gt=2$, $\beta=2$, (b) $gt$ with $\alpha=1$, $\beta=2$, (c) $\beta$ with $\alpha=1$, $gt=4$.}
\label{Q}
\end{figure}
\begin{figure}[htb]
\centering
\includegraphics[width=0.3\textwidth]{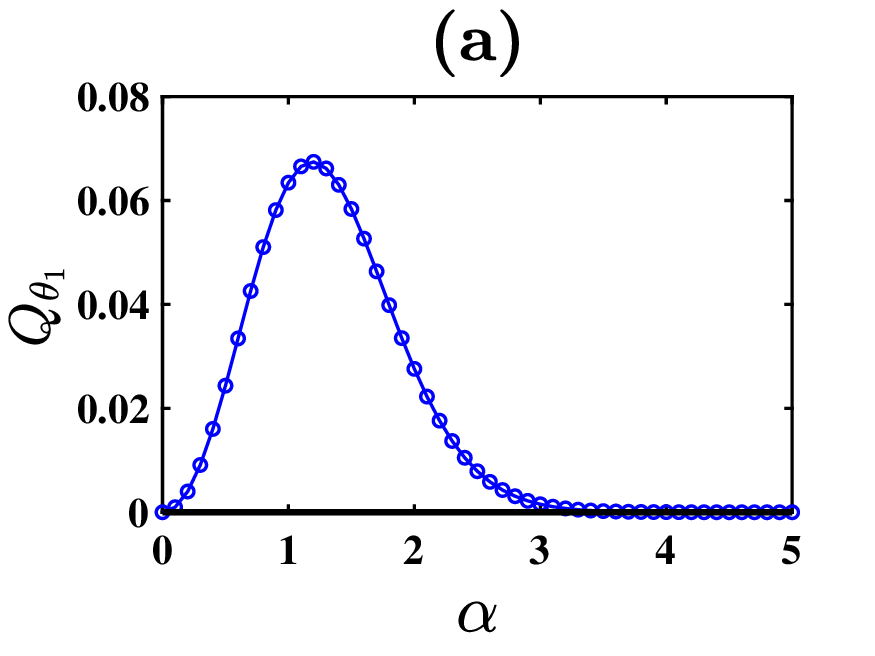}
\includegraphics[width=0.3\textwidth]{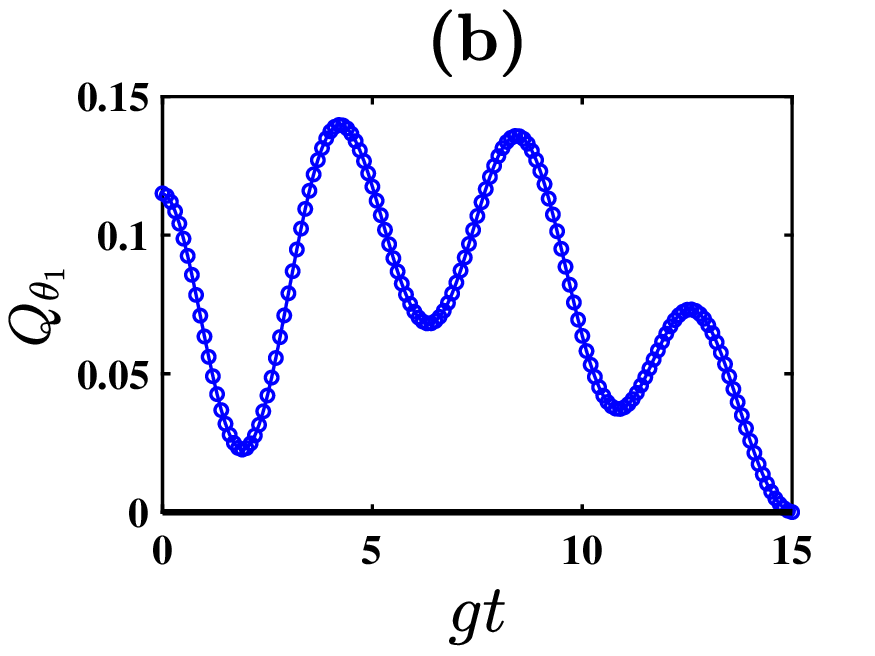}
\includegraphics[width=0.3\textwidth]{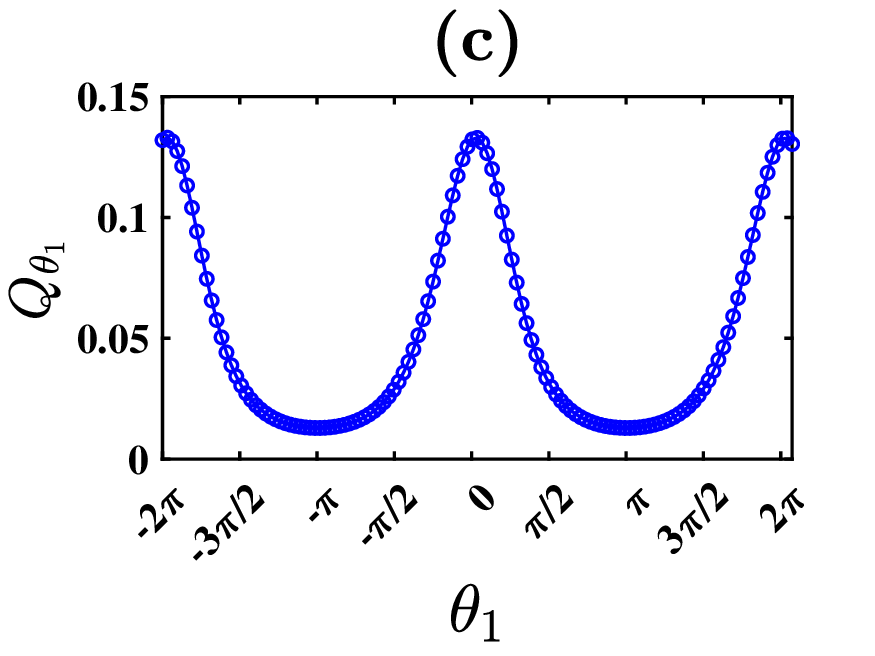}
\caption{Angular $Q$ is plotted as a function of (a) $\alpha$ with $gt=1$, $\theta_1=1$, (b) $gt$ with $\alpha=1$ and $\theta_1=1$, (c) $\theta_1$ with $\alpha=1$, $gt=1$.}
\label{angu}
\end{figure}
We have seen from Fig.~\ref{Q} that the behaviour of $Q$ function is similar with respect to $\alpha$ and $\beta$. The $Q$ parameter increases while $\alpha$ as well as $\beta$ increases, attains pick and then decreases to overlap with the line $Q=0$. The zero values of $Q$ imparts the nonclassical nature of the cavity field. One can clearly see that the $Q$ function has wave nature with reduced height of the waves as the scaled time $gt$ is increasing. Fig.~\ref{angu} shows the variation of the angular $Q$ function with displacement parameter $\alpha$, $gt$ and coherent state amplitude $\theta_1$. The pattern of the angular $Q_{\theta_1}$ plot is quite similar to that observed for the $Q$ parameter with respect to $\alpha$ and $gt$. That means increases with increasing displacement parameter $\alpha$, attains peak and then decreases to coincide with the line $Q_{\theta_1}=0$. Also, $Q_{\theta_1}$ shows wave nature with respect to $gt$. However, with increasing coherent state amplitude $\theta_1$, $Q_{\theta_1}$ behaves like wave which is in contrast of $Q$ function. The observed pattern shows the
relevance of studying both these distributions due to their
independent characteristics.

\subsection{Phase dispersion}
\begin{figure}[h]
\centering
\includegraphics[width=0.3\textwidth]{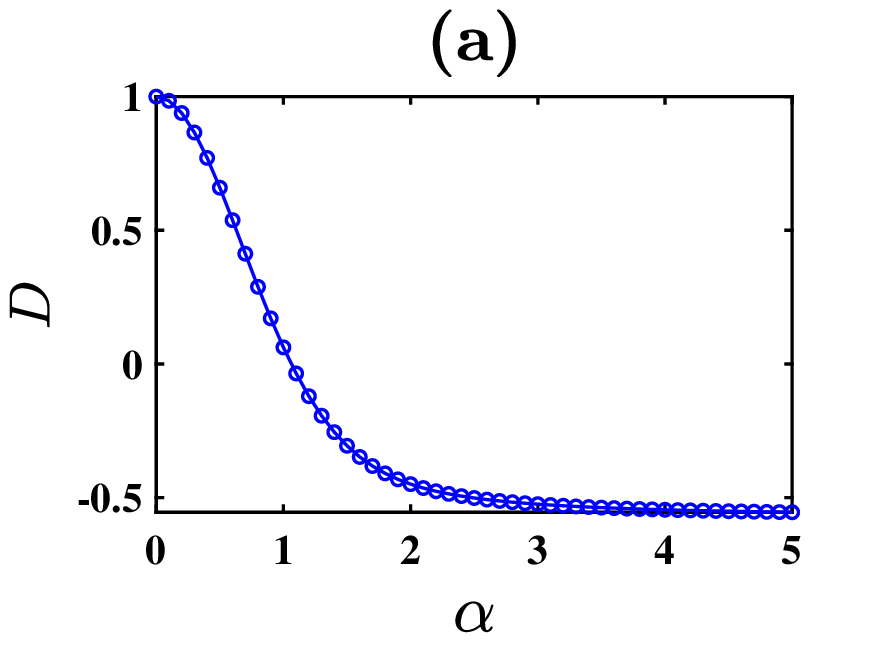}
\includegraphics[width=0.3\textwidth]{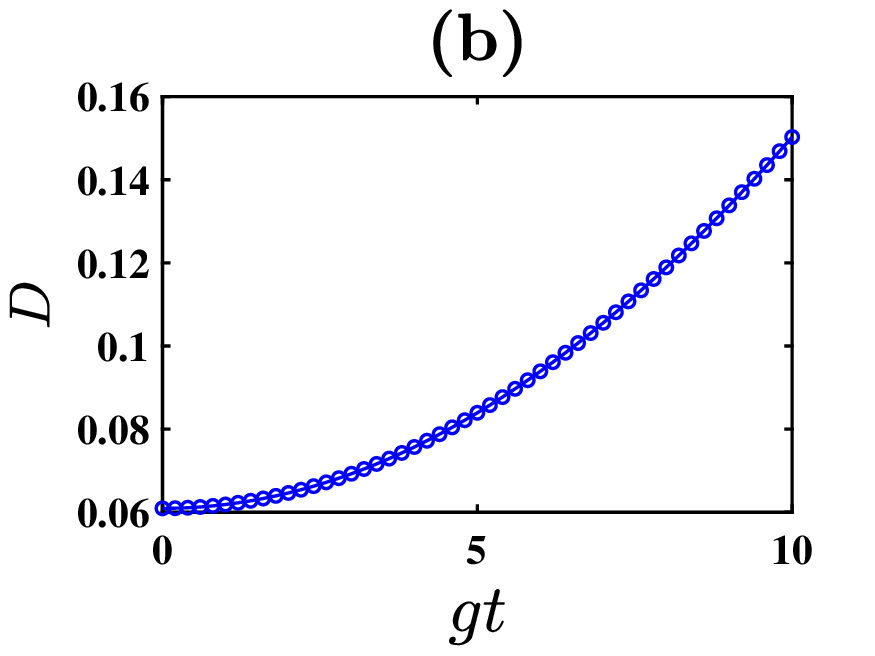}
\caption{Phase dispersion $D$ as a function of (a) $\alpha$ with $gt=2$, (b) $gt$ with $\alpha=1$.}
\label{D}
\end{figure}

A well-known measure quantum phase fluctuation by using the phase distribution function is phase dispersion. The variance, which has been used occasionally as a measure of phase fluctuation, has the disadvantage of being dependent on the origin of phase integration \cite{ban2}. In particular, the uniform phase distribution $P_\theta=1/2\pi$ corresponds to the maximum value of dispersion, i.e. 1. phase dispersion is a measure of phase fluctuation defined as \cite{k.thap}
\begin{equation}
\label{dispersion}
D=1-\left| \int_{-\pi}^{\pi}d\theta e^{-i\theta}P_\theta \right|^2
\end{equation}
For the cavity field state, the phase dispersion can be calculated as
\begin{align}
\nonumber
D&=1-\left| \int_{-\pi}^{\pi}d\theta e^{-i\theta}P_\theta \right|^2\\\nonumber
&=1-\frac{1}{2\pi}\left|\sum_{m,n=0}^{\infty}2\pi \delta_{m,\,n+1}\left({c^*_{a,m}+c^*_{b,m}}\right)\left({c_{a,n}+c_{b,n}}\right)\right|^2
\label{fdispersion}
\end{align}
where $\delta_{m,\,n+1}$ is the usual Dirac's delta function. The variation of phase dispersion is shown in Fig.~\ref{D}.

In Fig.~\ref{D}, it is observed that the phase dispersion $D$ exhibits a decreasing trend as the coherent parameter $\alpha$ increases. In contrast, with the increase in the value of scaled time $gt$, the quantum phase dispersion is found to increase. This suggests that the higher values of coherent parameter leads to smaller phase dispersion, whereas stronger interaction between the atom and the field results in higher phase dispersion. Moreover for $gt>2$, the phase dispersion as a function of $\alpha$ gives a broad curve.

\subsection{Antibunching}
Photon antibunching is one of the crucial evidence for the quantum nature of light. The second order correlation function $g^2(0)$ for any arbitrary single-mode radiation field $\ket\phi$ is defined as \cite{g2}\\
\begin{equation}
g^2(0)=\frac{\bra{\phi}a^{\dag 2}a^2\ket{\phi}}{\bra{\phi}a^{\dag}a\ket{\phi}^2}
\end{equation}
If a radiation field obeys $g^2(0)\geq 1$, the field is said to be bunched with super-Poissonian field statistics, whereas quantum states can violate this bound. Therefore a measurement of $g^2(0)<1$ necessarily signals the nonclassical characteristics of the field. As a consequence, the field is said to be antibunched with sub-Poissonian photon statistics.
\begin{figure}[h]
\centering
\includegraphics[width=0.3\textwidth]{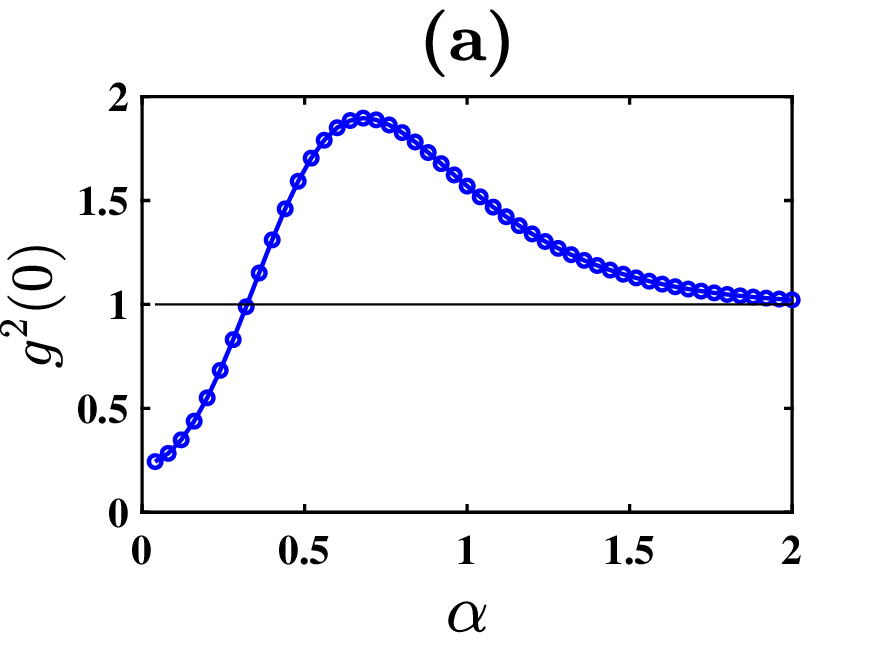}
\includegraphics[width=0.3\textwidth]{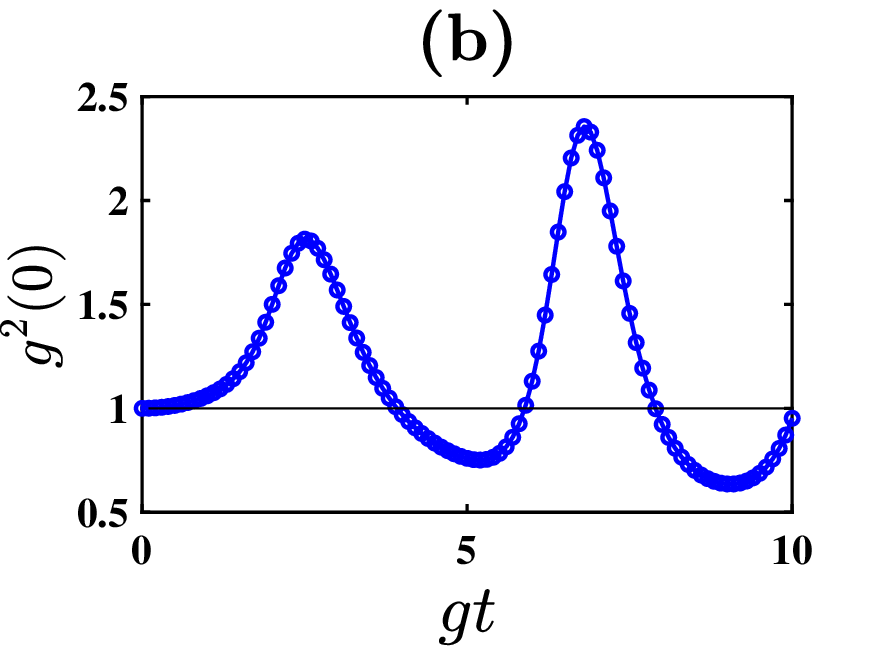}
\caption{Variation of second order correlation function $g^2(0)$ as a function of (a) $\alpha$ with $gt=3$, (b) $gt$ with $\alpha=1$.}
\label{g2}
\end{figure}
We have seen from Fig.~\ref{g2} that $g^2(0)$ corresponding to the cavity field is less then one for some values of $\alpha$ as well as $gt$ which imparts that the cavity field is antibunched. We have further observed that the cavity QED field exhibits antibunching behaviour for a short period of time that means the system presents signatures of temporary antibunching. In many experiments, it is important to stabilize the antibunching effect over time to ensure that the nonclassical properties of the light source are maintained. This steady-state antibunching can be achieved by using various techniques including feedback control and active stabilization \cite{wise}. However, there are also applications that may require temporary antibunching such as in the study of non-equilibrium quantum systems or in the generation of nonclassical light for specific experimental purposes \cite{mills}. One main usage is in quantum key distribution (QKD) where temporary antibunching can be used to generate secure cryptographic keys. Temporary antibunching can also be
used to generate entangled photon pairs in a random sequence which is necessary for the success of teleportation protocols. Thus temporary antibunching present for a limited time period is expected to play an important role in quantum information science.

In Fig.~\ref{g2}, the phenomena of antibunching is more likely to occur when the intensity of the light source is reduced (nearly 1), that means when the light source is not very bright. In the context of initial coherent state, for smaller values of coherent state amplitude $\alpha$, the state becomes less classical and starts to exhibit more quantum-like properties.

\section{Conclusion}
\label{sec4}

In a nutshell, phase dispersion defines how the phase of a wave packet spreads over time or space due to the factors like dispersion or interactions with a medium, phase distribution indicates the statistical distribution of phases within a wave or ensemble of waves, offering insights into interference patterns and quantum behaviour while phase fluctuations refer to random variations in phase that can arise from noise, thermal effects, or quantum uncertainties, affecting the stability of a wave. These measures collectively provide a more complete picture of the phase properties of a given quantum state. Their combined results allow researchers to assess coherence, interference patterns, stability, and quantum effects, shedding light on the quantum system's behaviour and interactions with its environment.

In this article, we have focused on an interacting atom-cavity field system driven by a classical field. This proposal is experimentally close to reach with currently available cavity QED techniques \cite{R1, R2, R3, R4}. Both microwave and optical regimes may be utilized for implementation of the scheme in cases of both open and closed cavities \cite{leib,schmidt}. In case of open cavities, which are defined by their apertures, photons can escape through the openings, resulting in shorter photon lifetimes and broader resonances compared to closed cavities. Photons in open cavities can interact more strongly with the external environment, leading to faster decoherence and increased sensitivity to external perturbations. But closed cavities are structures with reflective boundaries that confine electromagnetic waves, allowing the photons to bounce between the cavity walls and to experience minimal interaction with the external environment, leading to long coherence times. The cavity field state is obtained by assuming that the field is initially in a coherent state and atom enters the cavity in the excited state $\ket{a}$. We have calculated quantum phase distribution of the cavity field, which is further used to obtain the phase fluctuation and phase dispersion. The angular $Q$ function is also estimated. We have studied phase fluctuation using three Carruthers and Nieto parameter, $U$ revealed the presence of antibunching in our quantum state of interest. The phase distribution $P_{\theta}$ is of wave nature with respect to the displacement parameter $\alpha$ with varying amplitude, and $P_{\theta}$ is found to exhibit symmetry along $\theta=2\pi$ with respect to $gt$. In addition to the phase distribution, the second-order correlation function $g^2(0)$, the angular $Q$ function, and quantum phase fluctuation parameter $U$ manifest the nonclassical nature of the considered cavity field state.
\begin{center}
\textbf{ACKNOWLEDGEMENT}
\end{center}
Naveen Kumar acknowledges the financial support from the Council of Scientific and Industrial Research, Govt. of India (Grant no. 09/1256(0004)/2019-EMR-I). 

\end{document}